\documentclass[runningheads,a4paper]{llncs}
\usepackage[T2A]{fontenc}
\usepackage[cp1251]{inputenc}
\usepackage{amssymb}
\usepackage{graphicx}
\usepackage{epsfig}

\newcommand{\keywords}[1]{\par\addvspace\baselineskip
\noindent\keywordname\enspace\ignorespaces#1}
\def\be {\begin {eqnarray}}
\def\bb {\begin {eqnarray}}
\def\ee {\end {eqnarray}}

\begin{document}
\mainmatter  % start of an individual contribution

\title{
Symbolic-Numeric Algorithms for Computer Analysis of
  Spheroidal Quantum Dot Models }
\titlerunning{Symbolic-Numeric Algorithms for %Computer Analysis of
  Spheroidal Quantum Dot Models
}
\author{A.A. Gusev$^{1,2}$, O. Chuluunbaatar$^1$, V.P. Gerdt$^1$,
 V.A. Rostovtsev$^{1,2}$,\\ S.I. Vinitsky$^1$,
V.L. Derbov$^3$, V.V. Serov $^3$
}
\institute{
$^1$ Joint Institute for Nuclear Research, Dubna, Russia\\
$^2$ Dubna International University of Nature, Society $\&$Man, Dubna, Russia\\
$^3$ Saratov State University, Saratov, Russia\\
gooseff@jinr.ru}
\authorrunning{
A.A. Gusev et al.
}
\toctitle{Lecture Notes in Computer Science} \tocauthor{Authors'
Instructions} \maketitle
\begin{abstract}
A computation scheme for solving elliptic boundary value problems
with axially symmetric confining potentials
using different sets of one-parameter basis functions
is presented.
The efficiency of the proposed symbolic-numerical algorithms implemented in Maple
is shown by examples of spheroidal quantum dot models, for which energy spectra and eigenfunctions
versus the spheroid aspect ratio
were calculated within the
conventional effective mass approximation. Critical values of the aspect ratio,
at which the discrete spectrum of
models with finite-wall potentials is transformed into a continuous
one in strong dimensional quantization regime, were revealed  using the exact and
adiabatic classifications.

\end{abstract}

\section{Introduction}

To analyze the geometrical, spectral and optical
characteristics of quantum dots in the effective mass approximation
and in the regime of strong dimensional quantization following
\cite{Harrison},
many methods and models were used, including the exactly solvable model of a spherical
impermeable well  \cite{Gambaryan}, the adiabatic approximation for
a lens-shaped well confined to a narrow wetting layer~\cite{Hawrylak96} and
a hemispherical impermeable well  \cite{Hayk}, the model of strongly
oblate or prolate ellipsoidal  impermeable well \cite{79},  
as well as numerical solutions of the boundary value problems (BVPs) with separable variables
in the spheroidal coordinates for wells with infinite and finite wall heights
\cite{CNI2000,Trani,Lepadatu}. However, thorough comparative analysis
of spectral characteristics of models with different  potentials,
including those with non-separable variables, remains to be a challenging problem.
This situation stimulates the study of a wider class of model well potentials with application
of symbolic-numerical algorithms
(SNA) and problem-oriented software, developed by the authors
of the present paper during years~\cite{Prorammirovanie,casc07,kantbp,POTHMF,parobp,casc09}.

Here we analyse the spectral characteristics of the following models: a spherical quantum dot (SQD),
an oblate spheroidal quantum dot (OSQD) and a prolate spheroidal quantum dot (PSQD).
We make use of the Kantorovich method that reduces the problem to a set of ordinary differential
equations (ODE)
~\cite{put}. In contrast to the well-known method of adiabatic representation \cite{BhK58},
this method implies neither adiabatic separation of fast and slow variables, nor the presence of
a small parameter. We present a calculation scheme for solving
elliptical BVPs
with axially-symmetric potentials in cylindrical coordinates (CC),  spherical coordinates (SC),
oblate spheroidal coordinates (OSC), and  prolate spheroidal coordinates (PSC).
Basing on the
SNA
developed for axially-symmetric potentials, different sets of
solutions are constructed for the parametric
BVPs related to the fast subsystem, namely,
the eigenvalue problem solutions (the terms and the basis functions), depending upon
the slow variable as a parameter, as well as the matrix elements, i.e., the integrals of
the products of basis functions and their derivatives with respect to the parameter,
which are calculated analytically
by means of elaborated SNA MATRA,
implementing in MAPLE,
or  numerically using the program ODPEVP~ \cite{parobp},
implementing the finite-element method (FEM).
These terms and matrix elements form the matrices of variable coefficients in the set of
second-order ODE with respect to the slow variable. The
BVP for this set of ODE
is solved by means of the program KANTBP~\cite{kantbp},
also implementing the
FEM.
The efficiency of the calculation scheme and the SNA
used is demonstrated
by comparison of the spectra versus the  ellipticity of the prolate or oblate spheroid in the models of
 quantum dots with different confining potentials, such as the isotropic and anisotropic harmonic oscillator,
  the spherical and spheroidal well with finite or infinite walls, approximated
  by smooth short-range potentials, as well as by constructing the adiabatic classification of the states.

The paper is organized as follows. In Section \ref{1} the calculation scheme for solving elliptic
BVPs with axially-symmetric confining potentials is presented. In Section \ref{2} SNA
MATRA for solving parametric BVP and corresponding integrals implemented in Maple is described.
Section \ref{3} is devoted to the analysis
of the spectra of quantum dot models with three types of axially-symmetric potentials, including
the benchmark exactly solvable models. In Conclusion we summarize the results and discuss
the future applications of our calculation scheme and the SNA project
presented.

\section{The problem statement}\label{1}
Within the effective mass approximation under the conditions of
strong dimensional quantization the Schr\"{o}dinger equation
for the slow envelope of the wave function $\tilde\Psi(\tilde{\vec r})$ of a
charge carrier (electron $e$ or hole $h$) in the models of a
spherical, prolate or oblate spheroidal quantum dot (SQD, PSQD or
OSQD) has the form
\begin{eqnarray}
\label{sp01}
 \{\tilde {\hat H} -\tilde E \}\tilde \Psi(\tilde {\vec r}) =
 \{
({2\mu_p })^{-1} \tilde {\hat { {\vec P}}}^2 + \tilde U ( \tilde
{\vec r}) -\tilde E
 \}\tilde \Psi(\tilde {\vec r}) =0,
\end{eqnarray}
where $\tilde {\vec r}\in \bf R^{3}$ is the position vector of the particle having the effective mass $\mu_{p}=\mu_{e}$
(or $\mu_{p}=\mu_{h}$),
$\tilde {\hat {{\vec P}}}=-i\hbar \nabla
_{\tilde {\vec r}}$ is the momentum operator,
 $\tilde E$ is the energy of the particle $\tilde U ( \tilde {\vec r})$ is the axially-symmetric potential, confining
the particle motion in SQD, PSQD or OSQD.
In Model A   $\tilde U( \tilde {\vec r})$ is chosen to be the potential of isotropic or anisotropic
axially-symmetric harmonic oscillator with the angular frequency
$\tilde \omega =\gamma
_{\tilde r_0 } {\hbar }/({\mu_p \tilde r_0^2 })$, $\gamma _{\tilde
r_0 }\sim \pi^2/3$ being an adjustable parameter:
\begin{eqnarray}
\label{sp02}
\tilde U^{\mbox{А}} (\tilde {\vec r}) =
{\mu_p \tilde \omega ^2( \zeta_1(\tilde x^2+\tilde y^2)+\zeta_3\tilde z^2)}/{2},
\end{eqnarray}
$r_0 =\sqrt{\zeta_1(\tilde x_0^2+\tilde y_0^2)+\zeta_3\tilde z_0^2}$ is the radius of a spherical QD ($\zeta_1=1$,
$\zeta_3=1$) or that of a spheroidal QD ($\zeta_1= (\tilde r_0/\tilde a)^4$,
$\zeta_3=(\tilde r_0/\tilde c)^4$),
inscribed into a spherical one, where $\tilde a$ and $\tilde c$ are the semiaxes of the ellipse
which transforms into a sphere at $\tilde a=\tilde c=\tilde r_0$.
For Model B   $\tilde U (\tilde {\vec r})$ is the potential of a spherical or axially-symmetric well
\begin{eqnarray}
\label{sp03}
\tilde U^{\mbox{B}} (\tilde {\vec r})
=\{0,0 \le({\tilde x^2+\tilde y^2})/{\tilde a^2}+{\tilde z^2}/{\tilde c^2} < 1 ; \tilde U_0,
({\tilde x^2+\tilde y^2})/{\tilde a^2}+{\tilde z^2}/{\tilde c^2} \ge 1 \},
\end{eqnarray}
with  walls of finite or infinite height $1\ll \tilde U_0 < \infty$. For
Model C   $\tilde U ( \tilde {\vec r})$  is taken to be a spherical or
axially-symmetric diffuse potential
\begin{eqnarray}
\label{sp03VS}
\tilde U^{\mbox{C}} (\tilde {\vec r})
= \tilde U_0\left[1+\exp (({ (\tilde x^2+\tilde y^2)/\tilde a^2+\tilde z^2/\tilde c^2-1})/{s})\right]^{-1},
\end{eqnarray}
where $s$ is the edge diffusiveness parameter of the function, smoothly approximating
the vertical walls of finite height  $\tilde U_0$. Below we restrict
ourselves by considering Model B with infinite walls $\tilde U_0
\rightarrow\infty$ and Model C with walls of finite height $\tilde
U_0$. We
make use of the reduced atomic units: $a_B^*= {\kappa \hbar
^2}/{\mu _p e^2}$ is the reduced Bohr radius, $\kappa$ is the DC
permittivity,
 $E_R \equiv Ry^*= {\hbar ^2}/({2\mu _p {a_B^*}^2 })$
is the reduced Rydberg unit of energy, and the following dimensionless quantities are
introduced: $\tilde \Psi(\tilde{\vec r})= {a_B^*}^{-3/2}\Psi( {\vec r})$,
$2 {\hat H}= \tilde {\hat H}/{Ry^*}$, $2 {E}= {\tilde E}/{Ry^*}$,
$2 {U(\vec r)}= {\tilde U (\tilde {\vec r})}/{Ry^*}$, $\vec r=\tilde
{\vec r} /a_B^*$, $a=\tilde a/a_B^*$, $\tilde c=c/a_B^*$,
$r_0=\tilde r_0/a_B^*$, $\omega=\gamma_{r_{0}}/ r_0^{2}=\hbar
\tilde\omega/(2Ry^*)$.
For an electron with the reduced mass $\mu_p \equiv \mu_e = 0.067 m_0 $ at $\kappa =13.18$ in GaAs:
$a_B^*=102$\AA$=10.2$ nm, $Ry^*=E_R =5.2$ meV.

Since the Hamiltonian $\hat H$ in (\ref{sp01})--(\ref{sp03VS}) commutes with the $z$-parity operator
($z \to - z$  or $\eta  \to -\eta $), the solutions are divided into
even ($\sigma = + 1$) and odd ($\sigma = - 1$) ones.
The solution of Eq. (\ref{sp01}),  periodical with respect to the azimuthal angle
$\varphi$,  is sought in the form of a product
$\Psi(x_{f},x_{s},\varphi) = \Psi^{m\sigma} (x_{f},x_{s}
){e^{im\varphi }}/{\sqrt {2\pi } }$, where $m = 0,\pm 1,\pm 2,...$
 is the magnetic quantum number.
Then the function
$\Psi^{m\sigma} (x_{f},x_{s} )$
satisfies the following equation in the two-dimensional domain
$\Omega=\Omega_{x_f}(x_s)\cup\Omega_{x_s}\subset {\bf R}^2\backslash
\{0\}$, $\Omega_{x_f}(x_s)=( x_f^{\min }(x_{s}),x_f^{\max
}(x_{s}))$, $\Omega_{x_s}=( x_s^{\min },x_s^{\max })$:
\begin{eqnarray} &&
\left( \hat H_1 (x_{f};x_s) + \hat H_2 (x_{s})+ V (x_{f},x_{s}) - 2
{E}\right) \Psi^{m\sigma} (x_{f},x_{s} ) = 0. \label{sp09}
\end{eqnarray}
The Hamiltonian of the slow subsystem $\hat H_2(x_s)$ is expressed as
\begin{eqnarray} &&\label{sp09xs}
 \hat H_2(x_s)= \check H_2(x_s)
 = - \frac{1}{g_{1s}(x_{s})}\frac{\partial }{\partial x_{s}}g_{2s}(x_{s})
 \frac{\partial }{\partial x_{s} }+ \check V_{s}(x_{s}),
\end{eqnarray}
and the Hamiltonian of the fast subsystem
$ \hat {H}_1 (x_f ;x_s )$
is expressed via the reduced Hamiltonian  $\check H_f (x_f ;x_s)$
and the weighting factor $g_{3s}(x_s)$:
\begin{eqnarray}\label{sp09xf}&&
  \hat H_1(x_{f};x_s)=g_{3s}^{-1} (x_s )\check H_f (x_f ;x_s),\\&&\nonumber
 \check H_f (x_f ;x_s)= - \frac{1}{g_{1f}(x_{f})}\frac{\partial }{\partial x_{f}}g_{2f}(x_{f})
 \frac{\partial }{\partial x_{f} }+ \check V_{f}(x_{f})
 +\check V_{fs}(x_{f},x_{s}).
\end{eqnarray}
Table \ref{ktpy} contains the values of conditionally fast $x_{f}$ and slow $x_{s}$
independent variables,
the coefficients $g_{1s}(x_s)$, $g_{2s}(x_s)$, $g_{3s}(x_s)$, $g_{1f}(x_f)$, $g_{2f}(x_f)$,
and the reduced potentials $\check V_{f}(x_{f})$, $\check V_{s}(x_{s})$, $\check V_{fs}(x_{f},x_{s})$,
entering Eqs. (\ref{sp09})--(\ref{sp09xf}) for SQD, OSQD and PSQD in cylindrical  ($\vec x=(z,\rho,\varphi)$), spherical
 ($\vec x=(r,\eta=\cos\theta,\varphi)$),
 and oblate/prolate spheroidal ($\vec x=(\xi,\eta, \varphi)$)
 coordinates \cite{stigun}.
\begin{table}[t]\caption{The values of conditionally fast $x_{f}$ and slow $x_{s}$
independent variables,
the coefficients $g_{is}(x_s)$, $g_{jf}(x_f)$
and the potentials $\check V_{f}(x_{f})$, $\check V_{s}(x_{s})$, $\check V_{fs}(x_{f},x_{s})$,
in Eqs.(\ref{sp09})--(\ref{sp09xf})
 for SQD, OSQD and PSQD in cylindrical (CC), spherical (SC) and
 oblate $\&$ prolate spheroidal ({OSC} $\&$ {PSC}) coordinates with
$(d/2)^{2}=\pm( a^2 - c^2)$, $+$ for OSC, $-$  for PSC.}\label{ktpy}
\parbox{0.68\textwidth}{
\begin{tabular}
{|c|c|c|c|c|} \hline & \multicolumn{2}{|c|}{CC} &
\multicolumn{1}{|c|}{SC} &{OSC} $\&${PSC} \\\hline
\hline & OSQD& PSQD& SQD & OSQD $\&$ PSQD \\
\hline $x_f $& $ z $& $ \rho$& $\eta$&$\eta$
 \\
\hline $x_s $& $ \rho $& $ z$&
$ r $&$\xi$\\
\hline $g_{1f} $&$1$ & $ \rho $&
$1$&$1$\\
\hline $g_{2f} $&$1$ & $ \rho $&
$1 - \eta ^2 $&$1 - \eta ^2 $\\
\hline $g_{1s}$ & $ \rho $ & $1$&
$ r^2 $ &$1$\\
\hline $g_{2s} $& $ \rho $& $1$&
$ r^2 $&$\xi^2\pm 1$\\
\hline $g_{3s} $& $1$& $1$&
$ r^2 $&$1$ \\
\hline ${\check V}_f(x_f)$& $ \omega^2\zeta_3 z^2$& $m^{2}/\rho^{2}+
\omega^2\zeta_1 \rho^2$& $
m^{2}/g_{2f} $&$
m^{2}/g_{2f} \pm (d/2)^{2}g_{2f}2E
$\\
\hline ${\check V}_s(x_s)$&$m^{2}/\rho^{2}+ \omega^2\zeta_1 \rho^2$
& $ \omega^2\zeta_3 z^2$& $0$&$
\mp m^{2}/g_{2s}- ((d/2)^{2}g_{2s} - 1)2E
$\\
\hline ${\check V}_{fs}(x_f,x_s)$& $0$&$0$&
$ {\check V}( r,\eta) $&$ {\check V}(\xi,\eta)$\\
\hline
\end{tabular}

}
\end{table}
In spherical coordinates the potential $\check{V}(r,\eta)$
in Table \ref{ktpy}, using the definitions (\ref{sp02}), (\ref{sp03VS})
in the reduced atomic units, for Model A is expressed as
$${\check V}(r,\eta)={2}r^2 V(r,\eta)
={\omega ^2r ^4( \zeta_1(1-\eta^2)+\zeta_3\eta^2)},$$
and for Model C as
$${\check V}(r,\eta)={2} r^2 V(r,\eta)
={2}r ^2U_0\left[1+\exp (({ r^2 (
(1-\eta^2)/a^2+\zeta_3\eta^2/c^2)-1})/{s})\right]^{-1},$$
both
having zero first derivatives in the vicinity of the origin $r=0$
(equlibrium point). For Model B the potentials
${\check V}_{fs}$ are zero, since the potential (\ref{sp03}) is reformulated
below in the form of boundary conditions with respect to the
variables $x_f$ and $x_s$. The solution of the problem
(\ref{sp09})--(\ref{sp09xf}) is sought in the form of Kantorovich
expansion \cite{put}
\begin{eqnarray}
  \Psi^{Em\sigma}_i( x_{f},x_{s})=\sum_{j=1}^{j_{\max}}
 \Phi^{m\sigma}_j(x_{f}; x_{s})\chi_{j}^{(m\sigma i)}(E,x_{s}),
 \label{sp15}
\end{eqnarray}
using as a set of trial functions the eigenfunctions
$\Phi^{m\sigma}_j(x_{f}; x_{s})$  of the Hamiltonian $\check {H}_f
(x_f ;x_s )$ from  (\ref{sp09xf}), i.e., the solutions of the
parametric BVP
\begin{equation}
\label{sp17} \left\{ \check {H}_f (x_f ;x_s ) -
\check {\lambda}_i (x_s) \right\} \Phi _i^{m\sigma} (x_f ;x_s ) = 0,
\end{equation}
in the interval $x_{f}\in
\Omega_{x_f}(x_s)$
depending on the conditionally slow variable $x_{s}\in\Omega_{x_s}$ as on a parameter.
These solutions obey the boundary conditions
\begin{equation}
\label{sp17a} \lim \limits_{x_f \to x_f^{t}(x_{s})}\!\! \left(\!\!
{N_f^{(m\sigma)}(x_{s}) g_{2f} (x_f)\frac{d\Phi _j^{m \sigma} (x_f
;x_s)}{dx_f } + D_f^{(m \sigma)}(x_{s}) \Phi _j^{m \sigma} (x_f;x_s
)}\!\!\right)\!\! =\!\! 0
\end{equation}
in the boundary points
$\{x_f^{\min }(x_{s}),x_f^{\max}(x_{s})\}=\partial\Omega_{x_f}(x_s)$,
of the interval $\Omega_{x_f}(x_s)$.
In Eq. (\ref{sp17a}), $N_f^{(m\sigma)}(x_{s})\equiv N_f^{(m\sigma)}$,
$D_f^{(m\sigma)}(x_{s})\equiv D_f^{(m\sigma)}$,
unless specially declared, are determined by the relations
$N_f^{(m\sigma)} = 1$, $D_f^{(m\sigma)} = 0$ at $m = 0$, $\sigma = + 1$ (or
at $\sigma =0$, i.e  without parity separation),
$N_f^{(m\sigma)} = 0$, $D_f^{(m\sigma)} = 1$ at $m = 0$, $\sigma = - 1$ or at $m \ne 0$.
The eigenfunctions satisfy the orthonormality condition with the weighting function
$g_{1f}(x_f)$
in the same interval
$x_{f}\in \Omega_{x_f}(x_s)$:
\begin{equation}
\label{sp19}
\left\langle \Phi _i^{m\sigma} \vert \Phi _j^{m\sigma} \right\rangle =
\int\nolimits_{x_f^{\min}(x_{s})}^{x_f^{\max}(x_{s})}\Phi _i^{m\sigma} (x_f ;x_s )
\Phi _j^{m\sigma} (x_f ;x_s )
 g_{1f}(x_f)dx_f=\delta_{ij}.
\end{equation}
Here
$\check \lambda_1 (x_s)< ... < \check \lambda_{j_{\max } }(x_s) <...$ is
the desired set of real eigenvalues. Corresponding set of potential
curves $ 2{E}_1 (x_s ) < ... < 2{E}_{j_{\max } } (x_s ) <...$ of
Eqs. (\ref{sp09xf}) is determined by
$2 {E}_{j} (x_s )=g_{3s}^{-1} (x_s )\check \lambda_j (x_s)$.
Note, that for OSC and PSC the desired set of real eigenvalues
$\check \lambda_j (x_s)$ depends on a combined parameter,
$x_{s}\rightarrow p^{2}=(d/2)^{2}2E$, the product of spectral $2E$ and geometrical
$(d/2)^{2}$ parameters of the problem (\ref{sp09}).
The solutions of the problem (\ref{sp17})--(\ref{sp19}) for Models A and B are calculated in
the analytical form, while for Model C this is done using the program  ODPEVP \cite{parobp}.

Substituting the expansion  (\ref{sp15}) into Eq. (\ref{sp09}) in consideration of
(\ref{sp17}) and (\ref{sp19}), we get a set of ODE for the slow subsystem with respect to
the unknown vector functions
$ {\mbox{\boldmath $\chi$}}^{(m\sigma i)}(x_s ,E) \equiv
{\mbox{\boldmath $\chi$}}^{(i)}(x_s ) = (\chi _1^{(i)} (x_s
),...,\chi _{j_{\max } }^{(i)} (x_s ))^T$:
\begin{eqnarray} &&
\label{sp23}
\biggl(\!
-\!\frac{1}{g_{1s}(x_{s})}{\rm {\bf I}}
\frac{d}{d{ x_{s}}}g_{2s}(x_{s})\frac{d}{d{ x_{s}}}
+2  {\bf E} ({ x_{s}})+{\rm {\bf I}}\check{V}_{s}(x_s)
- 2{\rm {\bf I}} {E}
\biggr){\mbox{\boldmath $\chi$}}^{(i)}({ x_{s}})=
\\&&
=\!-\!\biggl(\frac{g_{2s}(x_s)}{g_{1s}(x_s)} {\rm {\bf W}} ({ x_{s}})+
\frac{1}{g_{1s}(x_{s})}\frac{dg_{2s}(x_{s}){\rm {\bf Q}}({x_s})}{d{x_s}}
\!+\!\frac{g_{2s}(x_{s})}{g_{1s}(x_{s})}{\rm {\bf Q}}({ x_{s}})
\frac{d}{d x_{s}}
\biggr)
{\mbox{\boldmath $\chi$}}^{(i)}({ x_{s}}).\nonumber
\end{eqnarray}
Here $2 {\bf E} (x_s )=\mbox{diag}(g_{3s}^{-1} (x_s )\check \lambda_j
(x_s))$, ${\rm {\bf W}}( x_s)$, and ${\rm {\bf Q}}( x_s)$ are
matrices of the dimension $j_{{\max} }\times j_{{\max} }$,
\begin{eqnarray}
W_{ij}(x_{s})=W_{ji}( x_{s})=
\int\nolimits_{x_f^{\min}(x_{s})}^{x_f^{\max}(x_{s})}g_{1f}(x_{f})
\frac{\partial\Phi_{i}(x_{f}; x_{s})}{\partial x_{s}}
\frac{\partial\Phi_{j}(x_{f}; x_{s})}{\partial x_{s}}dx_{f},
\label{sp23a}  \\
Q_{ij}(x_{s})=-Q_{ji}( x_{s})=
-\int\nolimits_{x_f^{\min}(x_{s})}^{x_f^{\max}(x_{s})}g_{1f}(x_{f})
\Phi_{i}(x_{f}; x_{s})
\frac{\partial\Phi_{j}(x_{f}; x_{s})}{\partial x_{s}}dx_{f},
 \nonumber
\end{eqnarray}
calculated analytically for Model B and by means of the program ODPEVP \cite{parobp} for Model C. Note, that for Model A in SC and CS
and  Model B in OSC and PSC the variables $x_f$ and $x_{s}$ are
separated, so that the matrix elements
$W_{ij}(x_{s})=Q_{ij}(x_{s})\equiv 0$ are put
into the  r.h.s. of Eq. (\ref{sp23})  and $\check{V}_s(x_{s})$ are substituted
from Table \ref{ktpy}.
The discrete spectrum solutions
$2E:    2E_1 < 2E_2 < ... < 2E_t < ...$,
that obey the boundary conditions in the points
$x_s^t = \{x_s^{\min } ,x_s^{\max }\}=\partial\Omega_{x_s}$
bounding the interval $\Omega_{x_s}$:
\begin{equation}
\label{eq6sar}
 \lim \limits_{x_s \to x_s^t }
 \left( {N_s^{( {m \sigma})} g_{2s} (x_s)
 \frac{d{\mbox{\boldmath $\chi$}}^{({m \sigma}p)}(x_s )}{dx_s }
 + D_s^{({m \sigma})} {\mbox{\boldmath $\chi$}}^{({m \sigma}p)}(x_s )} \right) = 0,
 \end{equation}
where $N_s^{({m \sigma})} = 1$, $D_s^{({m \sigma})} = 0$
at $m = 0, \sigma = + 1$ (or
at $\sigma =0$, i.e  without parity separation), $N_s^{({m \sigma})} = 0$, $D_s^{({m \sigma})} = 1$
at $m = 0, \sigma = - 1$ or at $m \ne 0$,
and the orthonormality conditions
\begin{eqnarray}
   \int\nolimits_{x_{s}^{\min}}^{x_{s}^{\max}}
    ({\mbox{\boldmath $\chi$}}^{(i)}({x_{s}}))^T
    {\mbox{\boldmath $\chi$}}^{(j)}({ x_{s}})g_{1s}(x_{s}) dx_{s}
     = \delta_{ij},
     \label{sp23n}
\end{eqnarray}
are calculated by means of the program KANTBP  \cite{kantbp}. To
ensure the prescribed accuracy of calculation of the lower part of
the spectrum  discussed below with eight significant digits we used
$j_{\max}=16$ basis functions in the expansion  (\ref{sp15}) and the
discrete approximation of the desired solution by  Lagrange  finite
elements of the fourth order with respect to the grid pitch
$\Omega^{p}_{h^{s}({x_s})}=[x^s_{\min},x^s_{k}= x^s_{k-1}+h_{k}^{s},
x^s_{\max}]$.

\section {SNA MATRA for calculus of the BVP and integrals
}\label{2}

To calculate effective potentials of problem
(\ref{sp23})--(\ref{sp23n}) in each value ${x_s}=x^s_k$ of the FEM
grid $\Omega^{p}_{h^{s}({x_s})}=[x^s_{\min},
x^s_{\max}]$,  we consider a discrete representation of solutions
$\Phi({x_f};{x_s})\equiv\Phi^{m\sigma}({x_f};{x_s})$ of the problem
(\ref{sp17}) by means of the FEM on the grid,
$\Omega^{p}_{h^f({x_f})}=[x^f_{0}\!=\!x^f_{\min}(x_{s}),$
$x^f_{k}=x^f_{k-1}+h_{k}^{f}, x^f_{\bar n}\!=\!x^f_{\max}(x_{s})]$,
in a finite sum:
 \begin{eqnarray}\label{nnn}
 \Phi({x_f};{x_s})=\sum_{\mu=0}^{\bar np}
 \Phi_{\mu}^h({x_s})N_{\mu}^p({x_f}) =\sum_{k=1}^{\bar n}\sum_{r=0}^{p}
 \Phi_{r+p(k-1)}^h({x_s})N^p_{r+p(k-1)}({x_f}),
 \end{eqnarray}
where $N_{\mu}^p({x_f})$ are local functions and
$\Phi^h_{\mu}({x_s})$ are node values of $\Phi(x^f_\mu;{x_s})$. The
local functions $N_{\mu}^p({x_f})$ are piece-wise polynomial of the
given order $p$ equals one only in the node $x^f_{\mu}$ and equals
zero in all other nodes $x^f_{\nu}\neq x^f_{\mu}$ of the grid
$\Omega^{p}_{h^f({x_f})}$, i.e.,
$N_{\nu}^p(x^f_\mu)=\delta_{\nu\mu}$, $\mu,\nu=0,1,\ldots,\bar np$.
The coefficients $\Phi_{\nu}({x_s})$ are formally connected with
solution $\Phi(x^{fp}_{k,r};{x_s})$ in a node
$x^f_{\nu}=x^{fp}_{k,r}$, $k=1,\ldots,\bar n$, $r=0,\ldots,p$:
 \begin{eqnarray} \Phi^h_{\nu }({x_s})=
 \Phi^h_{r+p(k-1)}({x_s}) \approx \Phi(x^{fp}_{k,r};{x_s}),
 \quad x^{fp}_{k,r}=x^f_{k-1}+\frac{h_k^{f}}{p}r.\nonumber
 \end{eqnarray}
The theoretical estimate for the ${\bf H^0}$ norm between the
exact and numerical solution has the  order  of
 \begin{eqnarray}
 \vert  \check \lambda_j({x_s})  -\!  \check \lambda^h_j({x_s})\!  \vert \!\leq \!
 c_1 h^{2p} ,~
 \left \Vert \! \Phi_j(x_{f}; x_{s})  \! -  \mbox{\boldmath$\Phi$}^h_j({x_s})\right \Vert_0 \!\leq\!
 c_2 h^{p+1},\label{eq13}
 \end{eqnarray}
where   $h^{f} = \max_{1<j<\bar n}h_{j}^{f}$ is maximum step of grid and
constants $c_1>0$, $c_2>0$ do not depend on step $h^{f}$ \cite{r27}.
It has been shown that we have a possibility to
construct  schemes for solving the BVPs and integrals with high order
of accuracy comparable with the
computer one  in according with the following estimations~\cite{parobp}
\begin{eqnarray}
&&\left| \frac{\partial\check \lambda_j(x_{s})} {\partial x_{s}}-
  \frac{\partial\check \lambda ^h_j(x_{s})}{\partial x_{s}}\right| \leq c_3h^{2p}, \,
\left\| \frac{\partial\Phi_j(x_{f};x_{s})}{\partial x_{s}} -
\frac{\partial\mbox{\boldmath$\Phi$}^h_j(x_{s})} {\partial
x_{s}}\right\|_0 \leq c_4h^{p+1},\\&&
\left|Q_{ij}(x_{s})-Q_{ij}^h(x_{s})\right| \leq c_5h^{2p}, \,
\left|W_{ij}(x_{s})-W_{ij}^h(x_{s})\right| \leq c_6h^{2p},
\label{eq13aa}
\end{eqnarray}
where $h^{f}$ is the grid step, $p$ is the order of finite elements,
$i$, $j$ are the number of the corresponding solutions, and
constants $c_3$, $c_4$, $c_5$ and $c_6$ do not depend on step $h^{f}$.
Proof is straightforward following the scheme of
proof of estimations (\ref{eq13}) in accordance with
\cite{r27,shultz}. The verification of the above estimations are
examined by numerical analysis on condensed grids and by comparison with
examples of exact solvable models A and B.

Let us consider the reduction of
BVP (\ref{sp17}), (\ref{sp19}) on the interval
$\Delta:~x^f_{\min}(x_s)< {x_f}< x^f_{\max}(x_s)$ with boundary
conditions (\ref{sp17a}) in points $x^f_{\min}(x_s)$ and $x^f_{\max}(x_s)$
 rewriting in the form
 \begin{eqnarray}
 {\bf A}({x_s})\Phi_j({x_f};{x_s})=  \check \lambda_j(x_{s}){\bf B}({x_s})\Phi_j({x_f};{x_s}),
 \label{3231}
 \end{eqnarray}
where ${\bf A}({x_s})$ is differential operator and ${\bf B}({x_s})$
is multiplication operator are differentiable by parameter
$x_s\in\Omega_{x_s}$. Substituting expansion (\ref{nnn}) to
(\ref{3231}) and integration with respect to ${x_f}$ by parts in the
interval $\Delta=\cup_{k=1}^{\bar n}\Delta_k,$ we arrive to a system
of the linear algebraic equations
 \begin{eqnarray}
 {\bf a}_{\mu\nu}^p({x_s})\Phi_{j,\mu}^h({x_s})=
 \check \lambda ^h_j(x_{s}){\bf b}_{\mu\nu}^p({x_s})\Phi_{j,\mu}^h({x_s}),
 \label{3231mat}
 \end{eqnarray}
in framework of the briefly described FEM. Using $p$-order Lagrange
elements \cite{r27}, we present below an Algorithm 1 for
construction of algebraic problem  (\ref{3231mat}) by the FEM in the
form of conventional pseudocode. It MAPLE realization allow us show
explicitly recalculation of indices $\mu$, $\nu$ and test of
correspondent modules of parametric matrix problems,
derivatives of solutions by parameter and calculation of integrals.

{\bf Algorithm 1} Generation of parametric algebraic problems\\
[0mm]
\underline{\underline{\hspace{\textwidth}}}
\noindent {\bf Input:}\\
$\Delta=\cup_{k=1}^{\bar n}\Delta_k=[x^f_{\rm min}(x_{s}),x^f_{\rm
max}(x_{s})]$, is interval of changing of independent variable
${x_f}$,
that boundaries depending on parameter $x_{s}=x^s_{k'}$; \\
 $h_k^{f}=x^f_k-x^f_{k-1}$
 is a grid step;\\
$\bar n$
is a number of subintervals $\Delta_k= [x^f_{k-1},x^f_k]$;\\
$p$ is a order of finite elements;\\
${\bf A}({x_s}),{\bf B}({x_s})$
are differential operators in Eq. (\ref{3231});\\
{\bf Output:}\\
$N_\mu^p(x_{f})$
is a basis functions in (\ref{nnn});\\
${\bf a}_{\mu\nu}^p({x_s}),~{\bf b}_{\mu\nu}^p({x_s})$
are matrix elements in system of algebraic equations (\ref{3231mat});\\
{\bf Local:} \\
$x^{fp}_{k,r}$ are nodes;
 $\phi^p_{k,r}({x_f})$
are Lagrange elements;
$\mu,\nu=0,1,\ldots,\bar np$ ;\\
[0mm]\underline{\underline{\hspace{\textwidth}}}\\
1:   for~$k$:=1~to~$\bar n$~do \\
\phantom{3:}\phantom{3}  $\quad$  for~$r$:=0~to~$p$~do \\
\phantom{3:}\phantom{3}  $\quad \quad x^{fp}_{k,r}=x^f_{k-1}+\frac{h_k^{f}}{p}r$\\
\phantom{3:}\phantom{3}  $\quad$  end~for; \\
\phantom{3:}\phantom{3}  end~for; \\
2:\phantom{3}$ \phi^p_{k,r}({x_f})=
%\frac
\prod_{r'\neq r}[({x_f}-x^{fp}_{k,r'})(x_{k,r}^{fp}-x^{fp}_{k,r'})^{-1}]$\\
3:\phantom{3}$ N_0^p({x_f}):=\{\phi^p_{1,0}({x_f}),{x_f}
\in \Delta_1;0,{x_f}\not\in \Delta_1 \};$\\
\phantom{3:}\phantom{3}  for~$k$:=1~to~$\bar n$~do \\
\phantom{3:}\phantom{3}  $\quad$  for~$r$:=1~to~$p-1$~do \\
\phantom{3:}\phantom{3}  $\quad \quad N_{r+p(k-1)}^p({x_f}):
=\{\phi^p_{k,r}({x_f}),{x_f} \in \Delta_k; 0, {x_f} \not\in \Delta_k,\}$\\
\phantom{3:}\phantom{3}  $\quad$ end~for; \\
\phantom{3:}\phantom{3}  $\quad
N_{kp}^p({x_f}):=\{\phi^p_{k,p}({x_f}), {x_f} \in \Delta_k;
                    \phi^p_{k+1,0}({x_f}),{x_f}\in\Delta_{k+1};
                    0,{x_f} \not\in \Delta_k\bigcup\Delta_{k+1}\};$\\
\phantom{3:}\phantom{3}  end~for; \\
\phantom{3:}\phantom{3} $ N_{\bar np}^p({x_f}):=\{ \phi^p_{\bar
n,p}({x_f}), {x_f} \in \Delta_{\bar n}; 0,
{x_f} \not\in \Delta_{\bar n}\};$ \\
4:\phantom{3}for~$\mu,\nu$:=0~to~$\bar np$~do \\
\phantom{3:}\phantom{3}$\quad$$\quad$ $ {\bf
a}_{\mu\nu}^p({x_s}):=\int\limits_\Delta  g_1(x_f) N_{\mu}^p({x_f})
{\bf A}({x_s})
N_{\nu}^p({x_f}) d{x_f}; $\\
\phantom{3:}\phantom{3}$\quad$$\quad$
$ {\bf b}_{\mu\nu}^p({x_s}):
=\int\limits_\Delta g_1(x_f) N_{\mu}^p({x_f}) {\bf B}({x_s}) N_{\nu}^p({x_f})    d{x_f};$\\
\phantom{3:}\phantom{3}  end~for;
\\[0mm]\underline{\underline{\hspace{\textwidth}}}\\

{\bf Remarks:}

1. For equation (\ref{sp17}) matrix elements of the operator (\ref{sp09xf}),
and $V({x_f};{x_s})=\check V_{fs}(x_f,x_s)+\check V_{f}(x_f)$
between local functions $N_{\mu}(x_f)$ and $N_{\nu}(x_f)$ defined in
same interval $\Delta_j$ calculated by formula, using
${x_f}=x^f_{k-1}+0.5h^f_k(1+\eta_f),$ $q,r=\overline{0,p}$:
$$\begin{array}l
\left({\bf a}({x_s})\right)_{\mu,\nu}= \int\limits_{-1}^{+1} \left
\{ {4\over (h^f_k)^2}g_{2f}(x_f)(\phi_{k,q}^p)^\prime
(\phi_{k,r}^p)^\prime +g_{1f}(x_f)V({x_f};{x_s})
\phi_{k,q}^p\phi_{k,r}^p \right
\} {h^f_k \over 2} d \eta_f,\\
\left({\bf b}({x_s})\right)_{\mu,\nu}
=\int\limits_{-1}^{+1}g_{1f}(x_f)\phi_{k,q}^p\phi_{k,r}^p
{h^f_k\over 2} d \eta_f,~~ \mu=q+p(k-1),~~ \nu=r+p(k-1).
\end{array}$$

2. If integrals do not calculated analytically, for example, see section \ref{3},
then they are calculated by numerical
methods \cite{r27}, by means of the Gauss quadrature formulae of the
order $p+1$.

3. For OSQD\&PSQD model C  the problem (\ref{sp17})--(\ref{sp19}) has been solved using
a grid
$\Omega^{p}_{h^f({x_f})}[x^f_{\min},x^f_{\max}]=-1(20)1$
(the number in parentheses denotes the number of finite elements
of order $p = 4$ in each interval).

Generally, 10-16 iterations are required for the subspace iterations to converge the
subspace to within the prescribe tolerance. If matrix ${\mathbf
a}^p\equiv{\bf a}^p({x_s})$ in Eq. (\ref{3231mat}) is not positively
defined, problem (\ref{3231mat}) is replaced by the following problem:
 \begin{eqnarray}
 \tilde{\mathbf a}^p  \> \mbox{\boldmath$\Phi$}^h =
  \tilde \lambda^h\> {\mathbf b}^p  \>
  \mbox{\boldmath$\Phi$}^h,\quad
 \tilde {\mathbf a}^p ={\mathbf a}^p -\alpha{\mathbf b}^p.\label{31b}
 \end{eqnarray}
The number $\alpha$ (the shift of the energy spectrum) is chosen
in such a way that matrix $\tilde {\mathbf a}^p$ is positive. The
eigenvector of problem (\ref{31b}) is the same, and
$\check\lambda ^h= \tilde \lambda ^h+\alpha$,
where shift $\alpha$ is evaluated by the Algorithm 2.

\textbf{Algorithm 2} {Evaluating the lower bound for the lowest eigenvalue of the
generalized eigenvalue problem}

In general case it is impossible to define the lower bound for the
lowest eigenvalue of Eq. (\ref{31b}), because the   eigenvalues
$\check \lambda ^h_{1}(x_s)<...<\check \lambda ^h_{i}(x_s)<...< \check \lambda ^h_{j_{max}}(x_s)$
is depended on the parameter $x_s$. But, we can
use the following algorithm to find the lower bound for the lowest
eigenvalue $\check \lambda ^h_{1}(x_s)$ at fixed value of $x_s$:

{\bf Step 1.} Calculate $\mathbf{ L\,D\,L^ T}$ factorization of
$\mathbf{A}^p-\alpha\mathbf{B}^p$.

{\bf Step 2.} If some elements of the diagonal matrix $\mathbf{D}$
are less than zero \\ \phantom{aaaaaaaaaaa} then put
$\alpha=\alpha-1$ and go to {\bf Step 3}, else go to {\bf Step 5}.

{\bf Step 3.} Calculate $\mathbf{ L\,D\,L^ T}$ factorization of
$\mathbf{A}^p-\alpha\mathbf{B}^p$.

{\bf Step 4.} If some elements of the diagonal matrix $\mathbf{D}$
are less than zero \\ \phantom{aaaaaaaaaaa} then put
$\alpha=\alpha-1$
and go to {\bf Step 3}, else put $\alpha=\alpha-0.5$\\
\phantom{aaaaaaaaaaa} and go to {\bf Step 8}.

{\bf Step 5.} Put $\alpha=\alpha+1$ and calculate
$\mathbf{L\,D\,L^ T}$ factorization of
$\mathbf{A}^p-\alpha\mathbf{B}^p$.

{\bf Step 6.} If all elements of the diagonal matrix $\mathbf{D}$
are greater than zero \\ \phantom{aaaaaaaaaaa} then put
$\alpha=\alpha+1$ and repeat {\bf Step 5}.

{\bf Step 7.} Put $\alpha=\alpha-1.5$.

{\bf Step 8.} End.

After using the above algorithm one should find the lower bound for
the lowest eigenvalue, and always
$\check \lambda ^h_{1}(x_s)-\alpha\leq 1.5$

\section{Spectral characteristics of spheroidal QDs}\label{3}
\begin{figure*}[t]
\includegraphics[width=0.44\textwidth,height=0.4\textwidth]{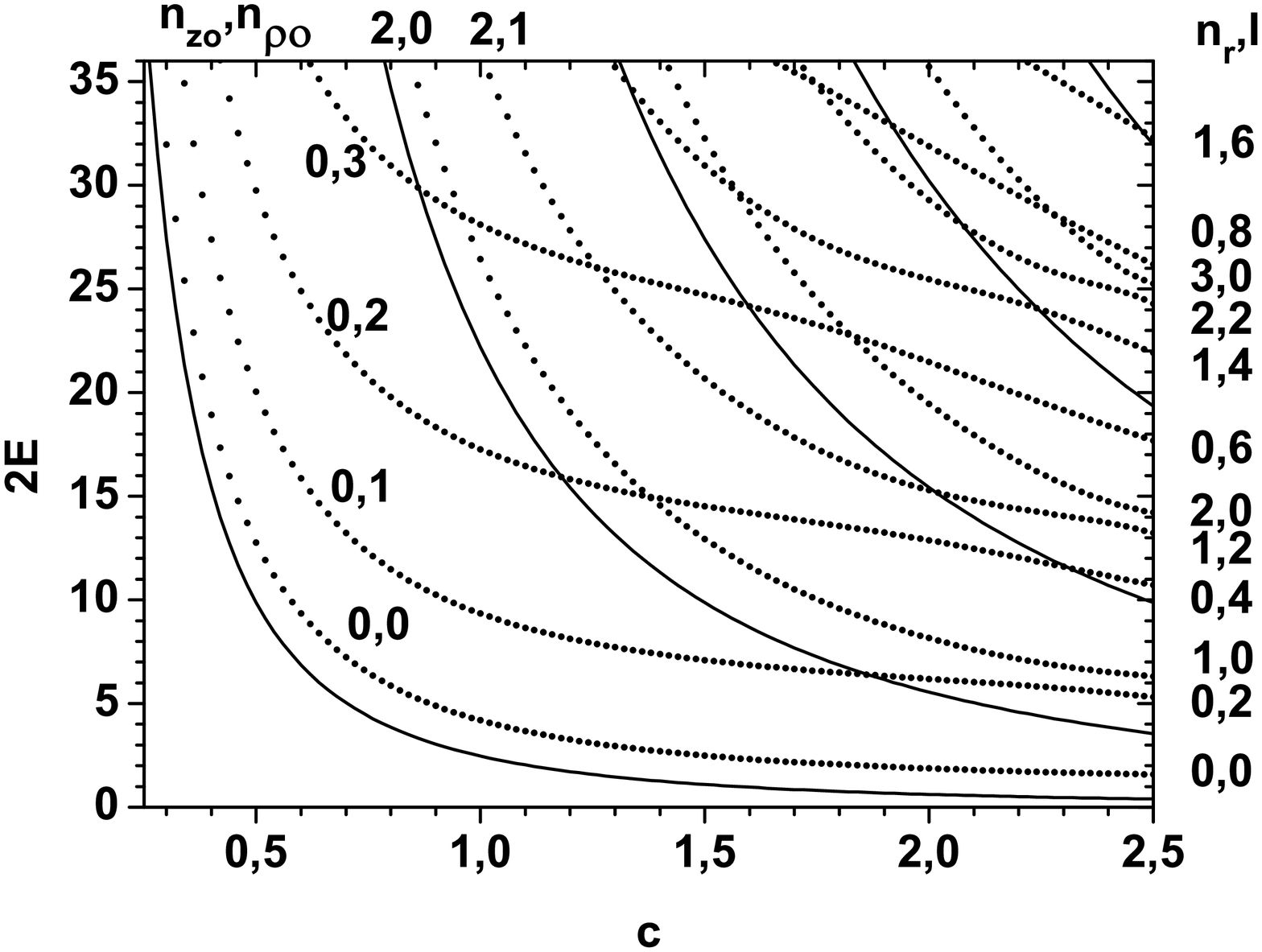}a) \hfill
\includegraphics[width=0.44\textwidth,height=0.4\textwidth]{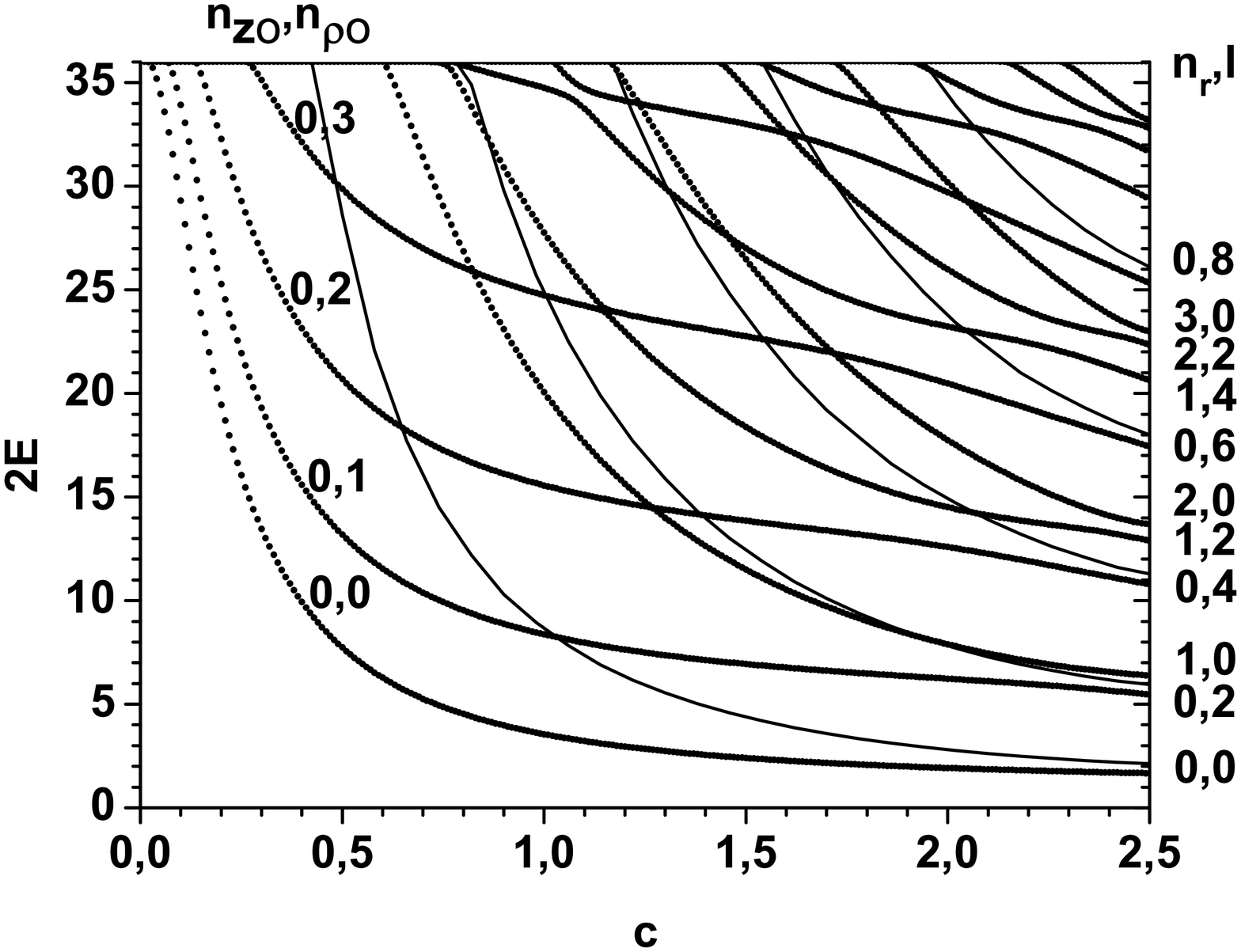}b)\\
\caption{The energies $2E=\tilde E/E_R$ of even $\sigma=+1$ lower states
for OSQD versus the minor
 $c$, $\zeta_{ca}=c/a\in(1/5,1)$ being the
spheroid aspect ratio: a) well with impermeable walls, b)
diffusion potential with   $2U_0=36$, $s=0.1$,  the major semiaxis
$a=2.5$ and $m=0$. Tine lines are minimal values
$2E_{i}^{min}\equiv 2E_{i}(x_{s}=0)$ of potential curves.} \label{enekts}
\end{figure*}
\begin{figure*}[t]
\includegraphics[width=0.98\textwidth]{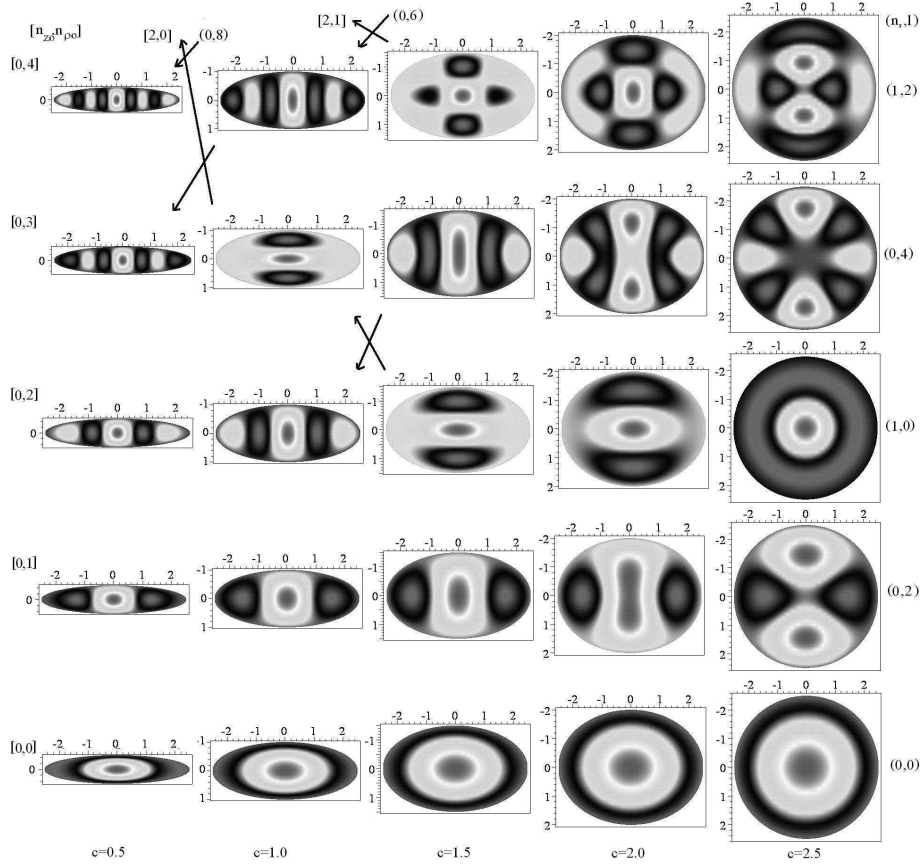}
\caption{Contour lines of the  first five even-parity wave functions $\sigma=+1$ in the $xz$ plane of Model B of OSQD
for  the major semiaxis $a=2.5$ and different values of the minor semiaxis $c$ ($\zeta_{ca}=c/a\in(1/5,1)$).
} \label{cwe}
\end{figure*}
\begin{figure*}[t]
\includegraphics[width=0.98\textwidth]{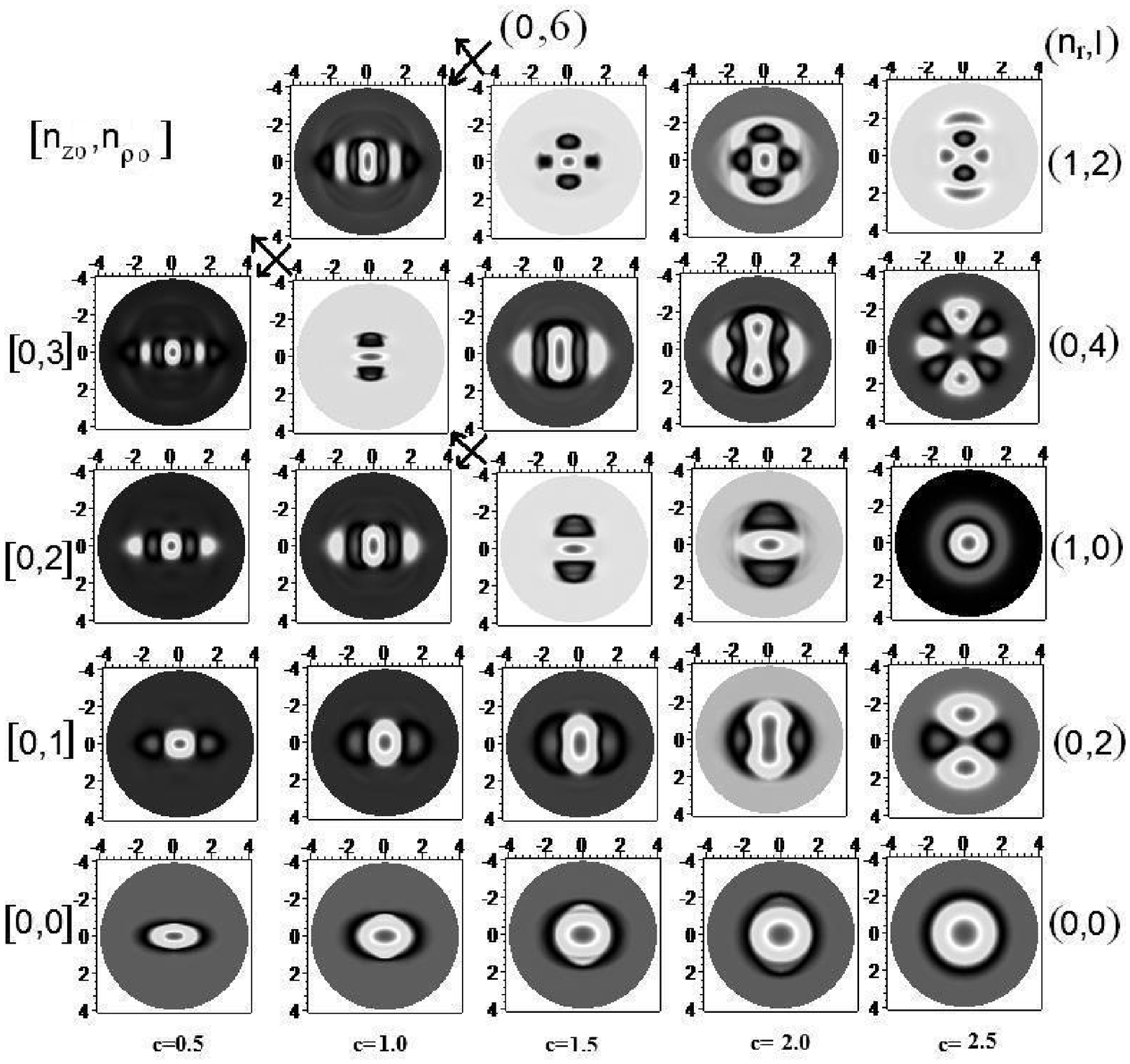}
\caption{Contour lines of the first five even-parity wave functions $\sigma=+1$ in the $xz$ plane of Model C of OSQD with
$2  U_0=36$ and $s=0.1$
for  the major semiaxis $a=2.5$ and different values of the minor semiaxis $c$ ($\zeta_{ca}=c/a\in(1/5,1)$).
} \label{cste}
\end{figure*}
\subsubsection{Models B and C for Oblate Spheroidal QD}
At fixed coordinate  $x_{s}$ of the slow subsystem the motion of the particle in the fast degree of
freedom $x_{f}$ is localized within the potential well having the effective width
\begin{equation}
\label{eq70}
\tilde L\left( x_s \right) = 2c\sqrt {1 - {x_s^2}/{a^2}} ,
\end{equation}
where $L=\tilde L/a_B^*$.
The parametric BVP (\ref{sp17})--(\ref{sp19})
at fixed values of the coordinate  $x_{s}$, $x_{s}\in (0,a)$,
is solved in the interval $x_f\in(- L\left( x_s \right)/2, L\left( x_s \right)/2)$
for Model C using the program ODPEVP, and for Model B the eigenvalues
$\tilde E_{n_{o}}\left( x_s \right)/E_R\equiv 2  E_{i}\left( x_s \right)$,
$n_{o}=i = 1,2,...$,
and the corresponding parametric eigenfunctions
$\Phi^\sigma_{i} \left( {x_f;x_s}\right)$,
obeying the boundary conditions (\ref{sp17a}) and the normalization condition  (\ref{sp19}),
are expressed in the analytical form:
\begin{equation}
\label{eq71}
\!\!\!\!\!\!\!\!2 E_{i}\left( x_s \right)\!=\!\frac{\pi ^2n_{o}^2}{
L^2\left( x_s \right)},\quad \Phi_{i}^{\sigma} \left(
{x_f;x_s}\right)\!=\!\sqrt{\frac{2}{ L\left( x_s \right)}}
\sin\left(\frac{\pi n_{o}}{2} \left(\frac{x_f}{ L\left( x_s
\right)/2}-1\right)\right),
\end{equation}
where the even solutions $\sigma=+1$  are labelled with odd $n_{o}=n_{zo}+1=2i-1,$
and the odd ones $\sigma=-1$
with even $n_{o}=n_{zo}+1=2i$, $i=1,2,3,...$ .
The effective potentials  (\ref{sp23a}) in Eq. (\ref{sp23}) for the slow subsystem are expressed analytically
via the integrals over the fast variable  $x_f$ of the basis functions (\ref{eq71})
and their derivatives with respect to the parameter $x_s$
{including states with both parities $\sigma=\pm1$}:
\begin{eqnarray}
&& 2  E_{i}(x_{s})=
\frac{a^2\pi^2n_{o}^2}{4c^2(a^2-x_s^2)},\quad
W_{ii}(x_{s})=\frac{3+\pi^2n_{o}^2}{12}\frac{x_s^2}{(a^2-x_s^2)^2},\label{sp23s}\\
&&W_{ij}(x_{s})=\frac{2n_{o}n_{o}'(n_{o}^2+n_{o}'{}^2)(1+(-1)^{n_{o}+n_{o}'})}{(n_{o}^2-n_{o}'{}^2)^2}\frac{x_s^2}{(a^2-x_s^2)^2},
\nonumber\\
&&Q_{ij}(x_{s})=\frac{n_{o}n_{o}'(1+(-1)^{n_{o}+n_{o}'})}{(n_{o}^2-n_{o}'{}^2)^2}\frac{x_s}{a^2-x_s^2},\quad n_{o}'\neq n_{o}.\nonumber
\end{eqnarray}

For Model B at  $c=a=r_{0}$ the OSQD turns into  SQD with known analytically expressed
energy levels $E_{t}\equiv E_{nlm}^{sp}$ and the corresponding eigenfunctions
\begin{equation}
2 E_{nlm}^{sp}\!=\!\frac{\alpha_{n_r+1,l+1/2}^2}{r_{0}^2},~
\Phi_{nlm}^{sp}(r,\theta,\varphi)
\!=\!\frac{\sqrt{2}J_{l+1/2}(\sqrt{2 E_{nlm}^{sp}}r)}
{r_{0}\sqrt{r}|J_{l+3/2}(\alpha_{n_r+1,l+1/2})|}
Y_{lm}(\theta,\varphi),
\end{equation}
where $\alpha_{n_r+1,l+1/2}$ are zeros of the Bessel function of
semi-integer index $l+1/2$, numbered in ascending order
$0<\alpha_{11}<\alpha_{12}< ...<\alpha_{iv}<...$ by the integer $i,v=1,2,3,...$. Otherwise one can use
equivalent pairs $iv\leftrightarrow\{n_r,l\}$ with
 $n_r=0,1,2,...$ numbering the zeros of Bessel function and  $l=0,1,2,...$
being  the orbital quantum number that determines the parity of
states $\hat{\sigma}=(-1)^l=(-1)^{m}\sigma$,
$\sigma=(-1)^{l-m}=\pm1$. At fixed  $l$  the energy levels  $\tilde
E_{nlm}/E_{R}=2 E_{t}$, degenerate with respect to the magnetic
quantum number $m$, are labelled with the quantum number
$n=n_r+1=i=1,2,3,...$ , in contrast to the spectrum of a spherical
oscillator, degenerate with respect to the quantum number
$\lambda=2n_{r}+l$. Figs. \ref{enekts}, \ref{cwe}, and \ref{cste}
show the lower part of non-equidistant spectrum $\tilde
E(\zeta_{ca})/E_{R}=2  E_t$ and the eigenfunctions
$\Psi^{m\sigma}_t$ from Eq. (\ref{sp15}) for even states OSQD Models
B and C at $m=0$. There is a one-to-one correspondence   rule
$n_{o}=n_{zo}+1=2n-(1+\sigma)/2, n=1,2,3,...$,
$n_\rho=(l-|m|-(1-\sigma)/2)/2$,  between the sets of spherical
quantum numbers  $(n,l,m,\hat\sigma)$ of SQD with radius  $r_0=a=c$
and spheroidal ones  $(n_\xi=n_r,n_\eta=l-|m|,m,\sigma)$ of OSQD
with the major $a$ and the minor $c$ semiaxes, and the adiabatic set
of cylindrical quantum numbers $(n_{zo},n_\rho,m,\sigma)$ at
continuous variation of the parameter $\zeta_{ca}=c/a$. The presence
of crossing points of the energy levels of similar parity under the
symmetry change from spherical $\zeta_{ca}=1$ to axial, i.e., under
the variation of the parameter $0<\zeta_{ca}<1$, in the BVP with two
variables at fixed  $m$ for Model B is caused by the possibility of
variable separation in the OSC \cite{stigun}, i.e. the r.h.s. of Eq.
(\ref{sp23}) equals zero. The transformation of eigenfunctions,
occurring in the course of a transition through the crossing points
in Fig. \ref{enekts}, is shown in Fig. \ref{cwe} for model B and in
Fig. \ref{cste} for model C (marked by arrows). One can see that
number nodes \cite{CurantGilbert} the eigenfunctions ordered in
according to increasing eigenvalues is not changed under the
transition from spherical to oblate spheroidal form. So, at small
value of deformation parameter ($\zeta_{ca}$ for OSQD or
$\zeta_{ac}$ for PSQD) there are nodes only along corresponding
major axis.
 For Model C at each value of the
parameter $a$ their is a finite number of discrete energy levels,
limited by the value $2U_0$ of the well walls height. As shown in
Fig. \ref{enekts}b, the number of levels of OSQD, equal to that of
SQD at $a=c=r_0$, is reduced with the decrease of the parameter  $c$
(or $\zeta_{ca}$), in contrast to Models A and B that have countable
spectra, and avoided crossings appear just below the threshold.
\begin{figure*}[t]
\includegraphics[width=0.44\textwidth,height=0.4\textwidth]{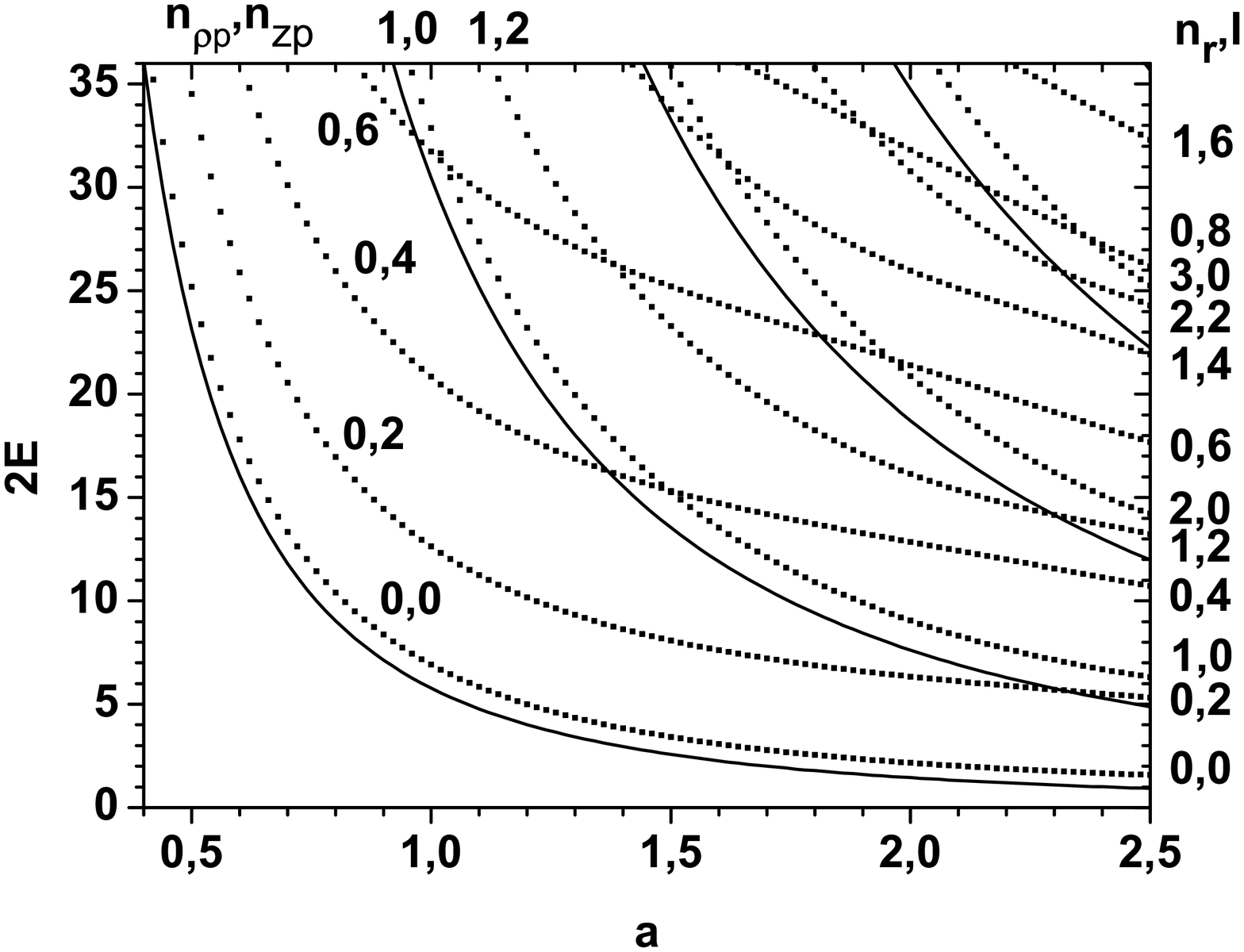} a) \hfill
\includegraphics[width=0.44\textwidth,height=0.4\textwidth]{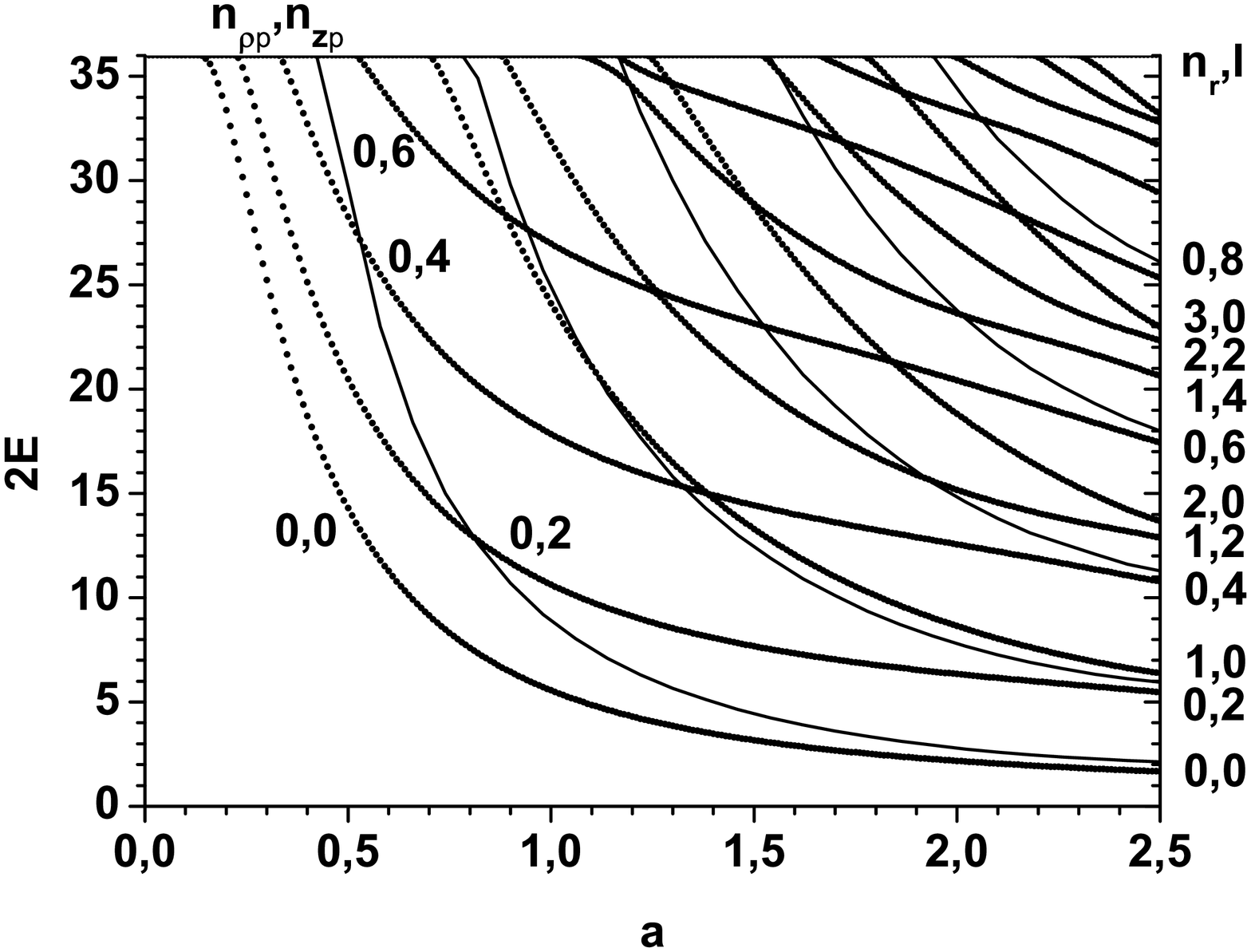}b)\\
\caption{The energies $2E=\tilde E/E_R$  of even $\sigma=+1$ lowest states
for PSQD depending on the minor semiaxis  $a$
$(\zeta_{ac}=a/c\in(1/5,1)$ is the spheroid aspect ratio): a) well
with impermeable walls, b)
diffusion potential,    $2U_0=36$, $s=0.1$, for the major semiaxis
$c=2.5$ and $m=0$. Tine lines are minimal values
$2E_{i}^{min}\equiv 2E_{i}(x_{s}=0)$ of potential curves.} \label{enektv}
\end{figure*}
\subsubsection{Models B and C for Prolate Spheroidal QD}
In contrast to OSQD, for PSQD at fixed coordinate  $x_s$
of the slow subsystem the motion of the particle is confined to a 2D potential well
with the effective variable radius
\begin{equation}
\label{eq99}
  \rho_{0} \left( x_{s} \right) = a\sqrt {1 - {x_s^2} / {c^2}} ,
\end{equation}
where $ \rho_{0}\left( x_{s} \right) = \tilde \rho_{0}\left( x_{s} \right)/ {a_B } $.
The parametric BVP
(\ref{sp17})--(\ref{sp19})
at fixed values of the coordinate $x_{s}$ from the interval $x_{s}\in (-c,c)$
is solved in the interval $x_f\in(0, \rho_{0}\left( x_{s} \right))$
for Model C using the program ODPEVP, while
for Model B the eigenvalues
$\tilde E_{n_{\rho p}+1}\left( x_s \right)/E_R\equiv
2E_{i}\left( x_s \right)$, $n_{\rho p}+1=i = 1,2,...$,
and the corresponding parametric basis functions
$\Phi^{m\sigma=0}_{i} \left( {x_f;x_s}\right)\equiv\Phi^{m}_{i} \left( {x_f;x_s}\right)$
without parity separation, obeying the boundary conditions
(\ref{sp17a}) and the normalization condition  (\ref{sp19}), are
expressed in the analytical form:
\begin{equation}
\label{eq100}
2 E _i \left( x_s \right) = \frac{\alpha _{n_{\rho p} + 1,|m|}^2}
 {  \rho_{0}^2\left( x_s \right)},\quad
\Phi^{m}_{n_{\rho p}}(x_{s})=\frac{\sqrt{2}}{  \rho_0\left( x_s
\right)} \frac{ J_{|m|}(\sqrt{2  E _{n_{\rho p}+1,|m|} \left( x_s
\right)}x_{f}) }{|J_{|m|+1}(\alpha_{n_{\rho p}+1,|m|})|},
\end{equation}
where $\alpha _{n_{\rho p}+ 1,|m|}=\bar J^{n_{\rho p}+ 1}_{|m|}$
are positive zeros of the Bessel function of the first kind  $J_{|m|}(x_f)$,
labeled in the ascending order with the quantum number $n_{\rho p}+1
=i= 1,2,...$.
The effective potentials (\ref{sp23a}) in Eq.(\ref{sp23}) for the
slow subsystem are calculated numerically in  quadratures via the
integrals over the fast variable  $x_f$ of the basis
functions(\ref{eq100}) and their derivatives with respect to the
parameter $x_s$ using SNA MATRA from Section 2.

Fig. \ref{enektv} illustrates the lower part of the non-equidistant
spectrum $E(\zeta_{ac})/E_{R}=2 \tilde E_t$ of even states of PSQD
Models B and C. There is a one-to-one correspondence rule  $n_{\rho
p}+1=n_p=i=n=n_r+1$, $i=1,2,...$ and $n_{z p}=l-|m|$ between the
sets of quantum numbers   $(n,l,m,\hat\sigma)$ of SQD with the
radius  $r_0=a=c$ and spheroidal ones
$(n_\xi=n_r,n_\eta=l-|m|,m,\sigma)$ of PSQD with the major $c$ and
the minor $a$ semiaxes, and the adiabatic set of quantum numbers
$(n=n_{\rho p}+1,n_{z p},m,\sigma)$ under the continuous variation
of the parameter $\zeta_{ac}=a/c$. The presence of crossing points
of similar-parity energy levels in Fig.   \ref{enektv} under the
change of symmetry from spherical $\zeta_{ac}=1$ to axial, i.e.,
under the variation of the parameter $0<\zeta_{ac}<1$, in the BVP
with two variables at fixed   $m$ for Model B is caused by the
possibility of variable separation in the PSC ~\cite{stigun}, i.e.
r.h.s. of Eq. (\ref{sp23}) equals zero. For Model C at each value of
the parameter $c$ there is also only a finite number of discrete
energy levels, limited by the value $2 U_0$ of the well walls
height. As shown in Fig. \ref{enekts}b the number of energy levels
of PSQD, equal to that of SQD at $a=c= r_0$, which is determined by
the product of mass $\mu_e$ of the particle, the well depth $\tilde
U_0$, and the square of the radius $ \tilde r_0$, is reduced with
the decrease of the parameter $\tilde a$ (or $\zeta_{ac}$) because
of the
promotion of the potential
curve (lower bound) into the continuous spectrum, in contrast to
Models A and B, having countable spectra. Note, that the spectrum of
Model C for PSQD or OSQD should approach that of Model B with the
growth of the walls height $U_0$ of the spheroidal well. However, at
critical values of the ellipsoid aspect ratio it is shown
that in the effective mass approximation both the terms (lower
bound) and the discrete energy eigenvalues in models of the B type
move into the continuum.
Therefore, when approaching the critical aspect
ratio values, it is necessary to use models such as
the lens-shaped self-assembled QDs with a quantum well confined
to a narrow wetting layer~\cite{Hawrylak96}
or if a minor semiaxis becomes comparable with the lattice constant
to consider models (see,e.g.\cite{Harper}), different from the effective mass approximation.

\section{Conclusion}
By examples of the analysis of energy spectra of SQD, PSQD, and OSQD models with thee types
of axially symmetric potentials, the efficiency of the developed computational scheme and
SNA is demonstrated. Only Model A (anisotropic harmonic oscillator potential)
is shown to have an equidistant spectrum, while Models B and C (wells with infinite and finite walls height)
 possess non-equidistant spectra. In Model C there is a finite number of energy levels.
 This number becomes smaller as the parameter $a$ or $c$ ($\zeta_{ac}$ or $\zeta_{ca}$) is reduced,
  because the potential curve (lower bound) moves into the continuum.
  Models A and B have countable discrete spectra. This difference in spectra allows
  verification of SQD, PSQD, and OSQD models using experimental data
  \cite{Gambaryan},
  e.g., photoabsorption, from which not only the energy level spacing,
  but also the mean geometric dimensions of QD may be derived \cite{79,Trani,Lepadatu}.
  It is
  shown that there are
critical values of the ellipsoid aspect ratio, at which in the
approximation of effective mass the discrete spectrum of models with
finite-wall potentials turns into a continuous one. Hence, using
experimental data, it is possible to verify different QD models
like the lens-shaped self-assembled QDs with a quantum well confined
to a narrow wetting layer~\cite{Hawrylak96},
or to determine the validity domain of the effective mass approximation,
if a minor semiaxis becomes comparable with the lattice constant
and to proceed opportunely to more adequate models, such as \cite{Harper}.

Note \textit{a posteriori}, that the diagonal approximation of the slow-variable ODE (\ref{sp23})
without the diagonal matrix element $W_{ii}$ (so called rude adiabatic approximation) provides
the lower estimate of the calculated energy levels. With this matrix element taken into account
(adiabatic approximation) the upper estimate of energy is provided, unless in the domain of the
energy level crossing points. Therefore, the Born-Oppenheimer (BO) approximation is, generally,
applicable only for estimating the ground state at an appropriate value of the small parameter.
For Model B in the first BO approximation $2 E_{i}\approx
E_i^{(0)}+E_{i}^{(1)}$ is given  by the minimal value of the slow
subsystem energy $E_1^{\min}(x_{s})$ in the equilibrium points
$x_s=0$ (namely, $E_i^{(0)}=\pi^2n_{o}^2/(2c)^2$ from Eq. (\ref{eq71}) for
OSQD and $E_i^{(0)}= \alpha_{n_{\rho p}+1}^2/a^2$  from Eq.
(\ref{eq100}) for PSQD), and by the corresponding energy values
$2 E_{i}^{(1)}=\pi(ac)^{-1} n_{o}(2n_{\rho}+|m|+1)$ and $E_{i}^{1}=  2(ac)^{-1}\alpha_{n_{\rho p}+1,|m|} (n_{z}+1/2)$ of the 2D and 1D
harmonic oscillator, respectively.
 In  \cite{Hayk} it is shown that the  terms $E_i(x_{s})$  allow high-precision
 approximation  by the Hulten potential. This can be accomplished by means of
 computer algebra software, e.g., Maple, Mathematica,  which allows
 (in the rude adiabatic approximation) to obtain the lower bound of the spectrum
 by solving transcendent equations, expressed analytically
 in terms of known special functions, and to use this approach for further development of
our SNA project.

The software package developed is applicable to the investigation of
impurity and exciton states in semiconductor nanostructure models.
Further development of the method and the software package is
planned for solving the quasi-2D and quasi-1D BVPs
with
both discrete and continuous spectrum, which are necessary for
calculating the optical transition rates, channeling and transport
characteristics in the models like quantum wells and quantum wires.

Authors thank Profs. K.G. Dvoyan, E.M. Kazaryan and H.A. Sarkisyan for
collaboration in the field and Profs. T. Sturm and C. Philips for support of
our SNA project. This work was done within the framework of the Protocol
No.3967-3-6-09/11 of collaboration between JINR and RAU in dynamics
of finite-dimensional quantum models and nanostructures in external fields.
The work was supported partially by RFBR (grants 10-01-00200 and
08-01-00604), and by the grant No. MK-2344.2010.2 of the President of
Russian Federation.

\end{document}